\documentclass[prl,aps,showpacs,twocolumn,preprintnumbers,amsmath,amssymb,superscriptaddress,longbibliography]{revtex4-1}
\usepackage[english]{babel}
\usepackage{amsmath,amssymb,bbm,mathrsfs,mathtools,multirow,diagbox,xcolor,%
  graphicx,tabularx,comment,amsfonts,dsfont,lettrine,units,comment}
\usepackage{graphicx}
\usepackage[colorlinks=True,linkcolor=red,citecolor=blue,urlcolor=blue]{hyperref}
\usepackage{bookmark}
\usepackage{xcolor}
\usepackage{chngcntr}
\usepackage{braket}
\usepackage{nicefrac}
\usepackage{bm}
\usepackage{bbm}
\usepackage{braket}
\usepackage{array}
\newcolumntype{C}{>{$}c<{$}}
\usepackage{newtxtext} 
\usepackage{ulem}

\newcommand{\diff}{\mathrm{d}}
\def\br{\mathbf{r}}
\def\bk{\mathbf{k}}

\def\ket#1{|#1\rangle }
\def\bra#1{\langle #1 |}

\usepackage{xstring} 

\newcommand{\customSection}[1]{{{\it{#1.}}---}}

\begin{document}

\title{Probing Topological Stability with Nonlocal Quantum Geometric Markers}

\author{Quentin Marsal}
 \affiliation{Department of Physics and Astronomy, Uppsala University, Box 516, 
751 20 Uppsala, Sweden}
\author{Hui Liu} \thanks{hui.liu@fysik.su.se}
 \affiliation{Department of Physics, Stockholm University, AlbaNova University Center, 106 91 Stockholm, Sweden}

 \author{Emil J. Bergholtz}
 \affiliation{Department of Physics, Stockholm University, AlbaNova University Center, 106 91 Stockholm, Sweden}
 
\author{Annica M. Black-Schaffer}
 \affiliation{Department of Physics and Astronomy, Uppsala University, Box 516, 
751 20 Uppsala, Sweden}

\begin{abstract}
    Spatially resolved local quantum geometric markers play a crucial role in the diagnosis of topological phases without long-range translational symmetry, including amorphous systems. Here, we focus on the nonlocality of such markers. We demonstrate that they behave as correlation functions independently of the material's structure, showing sharp variations in the vicinity of topological transitions and exhibiting a unique pattern in real space for each transition. Notably, we find that, even within the same Altland-Zirnbauer class, distinct topological transitions generate qualitatively different spatial signatures, enabling a refined, class-internal probe of topological stability.  
 As such, nonlocal quantum geometric indicators provide a more efficient and versatile tool to understand and predict the stability of topological phase transitions. 
\end{abstract}

\date{\today}%

\maketitle

One of the main interests in topological phases of matter, from both experimental and theoretical perspectives, lies in their robustness to disorder. 
Indeed, topology is insensitive to weak local perturbations as long as they do not change the symmetries of the system or close the bulk spectral gap~\cite{RMP_topological_classification}. 
However, for strong enough disorder, this topological protection can eventually break down, leading to transitions between topological and trivial phases~\cite{anderson1958, evers2008}. 
Such transitions often occur through the so-called pair levitation and annihilation mechanism ~\cite{Halperin_1982,laughlin_1984,Liu_2020}, with disorder closing the mobility gap.

While the framework of Anderson transitions provides a prototypical description of (de)localization in disordered topological crystals~\cite{evers2008}, it does not directly extend to amorphous materials---arguably the most common form of condensed matter, encompassing oxide glasses, disordered semiconductors and also photonic structures---where long-range translational symmetry is absent, but local order usually persists~\cite{zallen1998}. 
In such systems, there is no obvious energy scale analogous to the disorder strength used in ordinary lattice models, making it challenging to quantify how topology degrades, or possibly emerges, with structural randomness. 
Although topological phases in amorphous materials have been demonstrated theoretically~\cite{agarwala2017, marsal2020} and observed experimentally~\cite{mitchell2018, corbae2023}, their robustness remains poorly understood. 
Intriguingly, some topological phases appear to be more robust than others to amorphousness~\cite{hannukainen2022, andrei2020}, even within the same Altland-Zirnbauer (AZ) symmetry class \cite{altland1997} and for a given lattice realization. 
Thus, the manifestation and evolution of topology in amorphous matter obey principles yet to be uncovered.

Spatially resolved quantum geometric markers---such as the local Chern marker~\cite{bianco2011, loring2015, Cerjan_review2024} and its associated quantum metric~\cite{marzari1997, marrazzo2019, marsal2024}---have proven to be powerful tools for characterizing topology directly in real space.
Their locality enables the identification of topological phase transitions in crystals and amorphous systems alike, with their trace (or signature) having been used to reproduce the bulk topological invariant and, in finite systems, count the number of topological boundary states~\cite{Resta_2017}. 
At the same time, their nonlocality has been interpreted as a correlation function for electronic states, diverging at the associated topological phase transitions~\cite{molignini2023, desousa2023}.
However, despite these advances, these real-space markers still offer limited insight into how topology manifests and evolves in the presence of amorphousness.

In this work, we move beyond the current interpretation of quantum geometric markers and show that their nonlocal components, especially their characteristic real-space patterns, provides a direct probe of topological phase transitions and their stability in crystalline, disordered, and amorphous systems. 
In particular, the real-space patterns provide an intuitive understanding of why certain topological phases remain robust, while others collapse under amorphization. This resolves the puzzle of why different phases within the same symmetry class can exhibit markedly different stability, thereby providing a class-internal tool for phase differentiation. Further, by employing integrated markers, we establish the connection between real- and momentum-space formalisms.
This establish nonlocal quantum geometric markers as a key tool for understanding amorphous, and also crystalline, matter.

\customSection{Nonlocal quantum geometry}
The quantum geometric tensor $\mathcal{Q}^{n}$ of a given energy band (or set) $n$ measures the geometry of the electronic wave functions $|\psi_k^n\rangle$ in Hilbert space, usually defined in momentum space as~\cite{provost1980},
\begin{equation}
    \mathcal{Q}_{\mu\nu}^n = \frac{1}{\pi}\sum_{m\neq n}\int\diff k \braket{\partial_{k_\mu}\psi_k^n|\psi_k^m}\braket{\psi_k^m|\partial_{k_\nu}\psi_k^n},
\end{equation}
with momentum $k=(k_\mu,k_\nu)$.
 In two dimensions ($\mu,\nu=x,y$ throughout this work) the imaginary part of $\mathcal{Q}^{n}$ corresponds to the Chern number $C^n$~\cite{TKNN} through
\begin{equation}
    C^n = \frac{\mathcal{Q}_{xy}^n-\mathcal{Q}_{yx}^n}{2i}\label{eq:C},
\end{equation}
counting how many times the wave functions wrap the Hilbert space as momentum sweeps the Brillouin zone. 
The real part of $\mathcal{Q}^{n}$ is the quantum metric, which relates to the total volume of Hilbert space occupied by the given band. 
The trace of $\mathcal{Q}^{n}$ captures the gauge-independent contribution to the Wannier functions, thereby establishing a lower bound on the extent of any Wannier representation~\cite{vanderbilt_rmp},
\begin{equation}
    \Omega^n \equiv\text{tr}[\mathcal{Q}_{\mu\nu}^n]=\mathcal{Q}_{xx}^n+\mathcal{Q}_{yy}^n.\label{eq:omega}
\end{equation}
Alternatively, the quantum geometric tensor can be expressed in terms of position operators in real space~\cite{bianco2011},
\begin{equation}
    \mathcal{Q}_{\mu\nu}^n = \frac{4\pi}{A}\mathrm{Tr}[P\hat{r}_\mu Q\hat{r}_\nu P],\label{eq:QGT_r}
\end{equation}
where $\hat{r}_\mu$ denotes the position operator along $\mu$ direction, A is the area of the system, $P = \frac{A}{4\pi^2}\int\diff k \ket{\psi_k^n}\bra{\psi_k^n}$ is the projector on the targeted band $n$, and $Q = \mathbb{I}-P$. 
The trace is taken over the set of bands, {\it i.e.} over the eigenstates of the Hamiltonian.
In crystalline systems, the position-space and momentum-space formulations are equivalent in the thermodynamic limit~\cite{bianco2011}. 

The trace in Eq.~\eqref{eq:QGT_r} can be computed in any basis of the Hilbert space. Hence, for amorphous or disordered systems we can always use a basis of localized orbitals $\ket{r,\alpha}$, one for each lattice site, $r$ and orbital $\alpha$.
The elements of the matrix representation of the quantum geometric tensor then define a two-site correlator
\begin{equation}
    \mathcal{Q}_{\mu\nu}^n(r, r') = \frac{4\pi}{A}\sum_\alpha\bra{r,\alpha}P\hat{r}_\mu Q\hat{r}_\nu P\ket{r', \alpha}.
\end{equation}
For $r=r'$, this equation becomes Eq.~\eqref{eq:QGT_r}. For $r\neq r'$, we extend the Chern number and the conventional quantum metric to define a nonlocal Chern marker~\cite{molignini2023} 
\begin{equation}
    C^n(r, r') = -i\frac{2\pi}{A}\sum_\alpha\bra{r,\alpha}P\hat{r}_x Q\hat{r}_y P-P\hat{r}_y Q\hat{r}_x P\ket{r' ,\alpha},
\end{equation}
and a nonlocal quantum metric
\begin{equation}
    \Omega^n(r, r') = \frac{4\pi}{A}\sum_\alpha\bra{r, \alpha}P\hat{r}_x Q\hat{r}_x P+P\hat{r}_y Q\hat{r}_y P\ket{r', \alpha},
\end{equation}
respectively. 
In this work, we establish these observables as fundamental tools for assessing the stability of topological phases of crystalline, disordered, and amorphous systems. 

\customSection{Crystalline and amorphous systems}
To illustrate our general idea, we consider both crystalline and amorphous realizations of the same two-dimensional Hamiltonian in real space,
\begin{equation}
    H = \sum_i H_0c_i^\dagger c_i + \sum_{\left<i,j\right>}H_{ij}c_i^\dagger c_j\label{eq:H},
\end{equation}
with $H_0 = (M+4t)\sigma_z$, and 
\begin{equation}
    H_{ij} = t [i\cos(\theta_{ij})\sigma_x + i\sin(\theta_{ij})\sigma_y-\sigma_z],\label{eq:hopping}
\end{equation}
where $t$ is the nearest-neighbor coupling strength, $\theta_{ij}$ the angle between the bond direction and the $x$-axis, and $M$ the onsite potential.
Here hermiticity of $H$ imposes $H_{ij} = H_{ji}^\dagger$, which is fulfilled for $\theta_{ij} = \theta_{ji}+\pi$. 
A crystalline system is obtained by setting the Hamiltonian Eq.~\eqref{eq:H} on a square lattice.
Due to translation symmetry, we then get the Bloch Hamiltonian,
\begin{multline}
    \mathcal{H}(k) = [M-2t(\cos(k_x)+\cos(k_y)-2)]\sigma_z \\+ 2t\sin(k_x)\sigma_x + 2t\sin(k_y)\sigma_y\label{eq:Hk},
\end{multline}
which represents a two-band Chern insulator, closely related to the Qi-Wu-Zhang model~\cite{QWZ2006} and belonging to the A symmetry class of the AZ classification \cite{altland1997}. 
It exhibits multiple topological phase transitions as the mass term $M/t$ is tuned (see below), making it a convenient framework to demonstrate how the nonlocal quantum geometric tensor captures the stability of topological phases. To model disorder, we add a local disorder term, $H_0(i) = (M +\delta M\epsilon_i + 4t)\sigma_z$, where $\epsilon_i$ is a random number uniformly distributed in $[-1,1]$.
To obtain a fully amorphous lattice, still sharing a similar local environment, we first build an amorphous threefold coordinated lattice by starting from a random set of seeds, from which we draw the Voronoi tesselation, before finally mergeing all threefold-coordinated sites into pairs of neighbors, to produce a four-fold coordinated lattice~\cite{marsal2020}. 
For these systems, we first use the local Chern marker and the local quantum metric to identify the topological phases. 
We then introduce the nonlocal quantum geometric tensor in the vicinity of topological phase transitions to probe the topological stability.
Since the Hamiltonian~\eqref{eq:H} has only two bands, we can drop the band index $n$ and simply write $\Omega = \Omega^1 = \Omega^2$ and $C = C^1 = -C^2$.

\begin{figure}
    \centering
    \includegraphics[width=\linewidth]{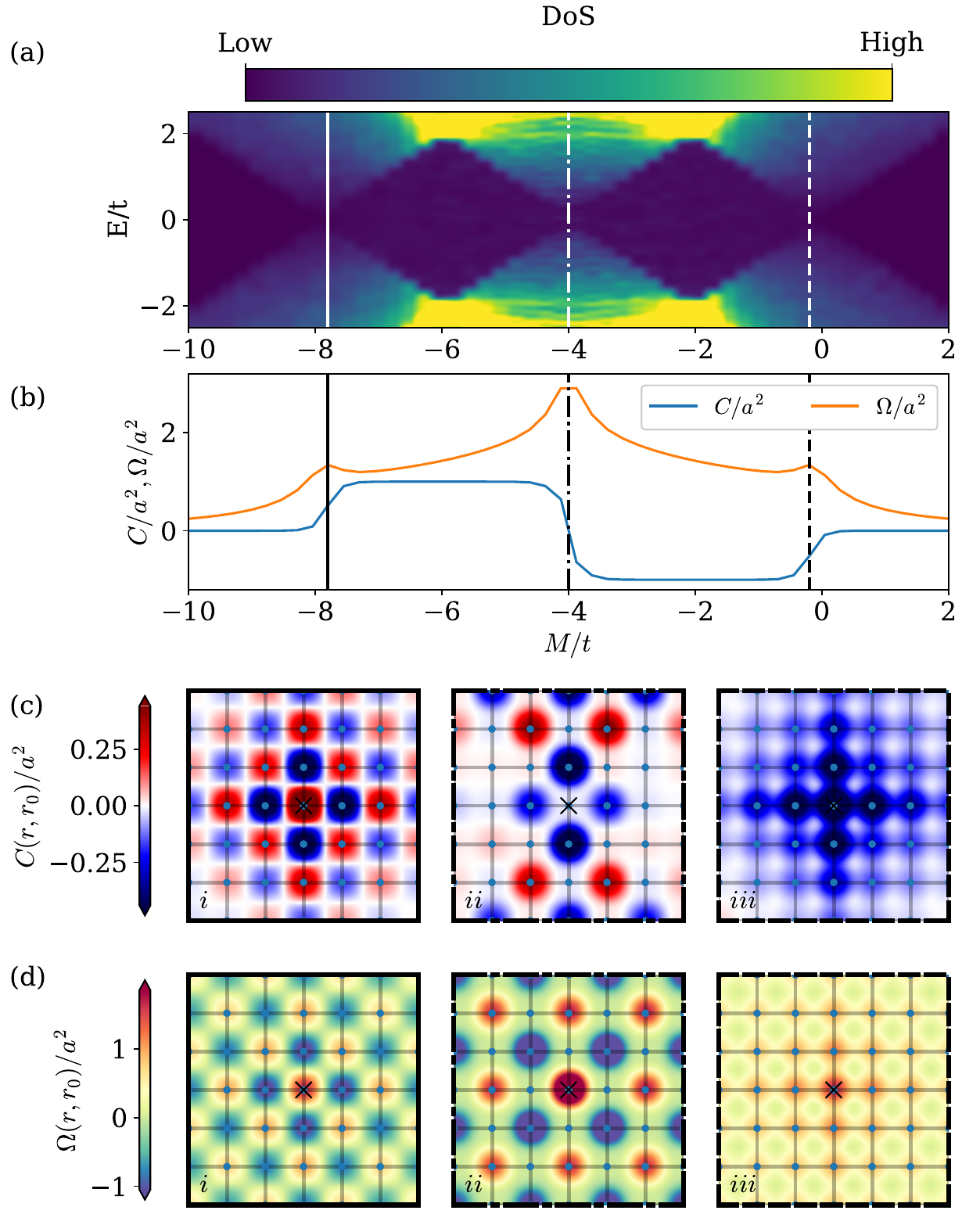}
    \caption{Nonlocal quantum geometry at the topological phase transition for the crystalline 
    system. (a) Energy spectrum as a function of $M/t$. Gap closings occur at $M/t = 0, -4, -8$, indicating topological phase transitions.
    (b) Chern number $C$ and quantum metric $\Omega$. The finite width of the phase transitions stem from the finite size of the sample.
    (c-d) Real-space patterns of the nonlocal Chern marker $C(r,r_0)$ (c) and quantum metric $\Omega(r,r_0)$ (d) close to each topological transition: ($i$) $M/t=-7.8$, ($ii$) $\ M/t = -4$, and ($iii$) $\ M/t = -0.2$. Solid, dashed, and dashed-dotted vertical lines and frame contours indicating the different $M$ values. Crystalline lattice is drawn for reference and black cross indicates reference site $r_0$.}
    \label{fig:crystal}
\end{figure}

\customSection{Topological phase transitions in a crystal}
We first study the behavior of nonlocal quantum geometry in the crystalline lattice.
The energy spectrum of the Hamiltonian Eq.~\eqref{eq:Hk} only depends on the ratio $M/t$.
As Fig.~\ref{fig:crystal}(a) shows, the density of states (DoS) exhibits three gap closings at $M/t = 0$, $-4$ and $-8$.
These gap closings could also be predicted analytically based on $\mathcal{H}(k)$: its gap can only close in the Brillouin zone where $\sin(k_x) = \sin(k_y) = 0$, and only if its $\sigma_z$-term vanishes at those points. 
At these points, the gap width $\Delta(k_x, k_y)$ is $\Delta(0,0) = 2|M|$, $\Delta(0,\pi)=\Delta (\pi,0) = 2|M+4t|$ and $\Delta(\pi,\pi) = 2|M+8t|$, hence the gap closing points in Fig.~\ref{fig:crystal}(a).

Each gap closing separates two different topological phases that can be diagnosed using the Chern number in Eq.~\eqref{eq:C}, see Fig.~\ref{fig:crystal}(b). Large onsite splitting leads to trivial phases, but for $|M+4t|<4t$ two different topological phases emerge with Chern numbers $\pm 1$.
The topological phase transitions can alternatively be seen in the divergence of the quantum metric $\Omega$, using Eq.~\eqref{eq:omega}.
Such peaks in the local quantum metric is expected based on its relation to localization of the eigenstates~\cite{marzari1997}: a topological phase transition requires closing the bulk spectral gap of the system, {\it i.e.} it delocalizes the electronic wave functions. 

The nonlocal part of the quantum geometric tensor close to each topological phase transition is shown in Fig~\ref{fig:crystal}(c-d).
There, we choose an arbitrary reference site in the bulk, denoted $r_0$ and indicated by a cross, and then plot $\Omega(r,r_0)$ and $C(r,r_0)$ as a function of $r$. Previously it has been shown that these take non-zero values for $|r-r_0|<\xi$, with $\xi$ diverging at the phase transition~\cite{chen2017, chen2019, molignini2023}.
Importantly, we here find that the pattern formed in position space by each part of the quantum geometry is unique to each topological phase transition: the nonlocal Chern marker and quantum metric decay monotonically away from $r_0$ in the vicinity of $M = 0$, but they acquire different modulations at both other transitions $M = -4t$, $8t$.
These patterns can be traced back to the momentum-space perspective. Since momentum and position spaces are related by a Fourier transform~\cite{molignini2023}, a peak at non-zero $k$ in momentum space maps to phase modulations in the position space and vice-versa. 
Similarly, a uniform spread in one space results in a sharp peak in the other.
Close to the topological transitions, the quantum geometry peaks in momentum space around the momentum where the gap closes. 
This leads to a broader spread of the real-space quantum geometry, which also gets modulated with the corresponding momentum. In particular, at $M = 0$, the gap closes at $\bk = (0,0)$, so the quantum geometry shows no oscillations (Fig.~\ref{fig:crystal}(c-d,$iii$), with dashed frame contour). 
At $M = -4t$, the quantum geometry acquires two superimposed modulations, $\bk = (0,\pi/a)$ and $(\pi/a,0)$, where $a$ is the lattice spacing, resulting in a cross-squared-shaped quantum geometry (Figure~\ref{fig:crystal}(c-d,$ii$)).
At $M = -8t$, the gap closes at $\bk = (\pi/a, \pi/a)$, hence the checkerboard pattern quantum geometry (Figure~\ref{fig:crystal}(c-d,$i$)) . 

\begin{figure}
    \centering
    \includegraphics[width=\linewidth]{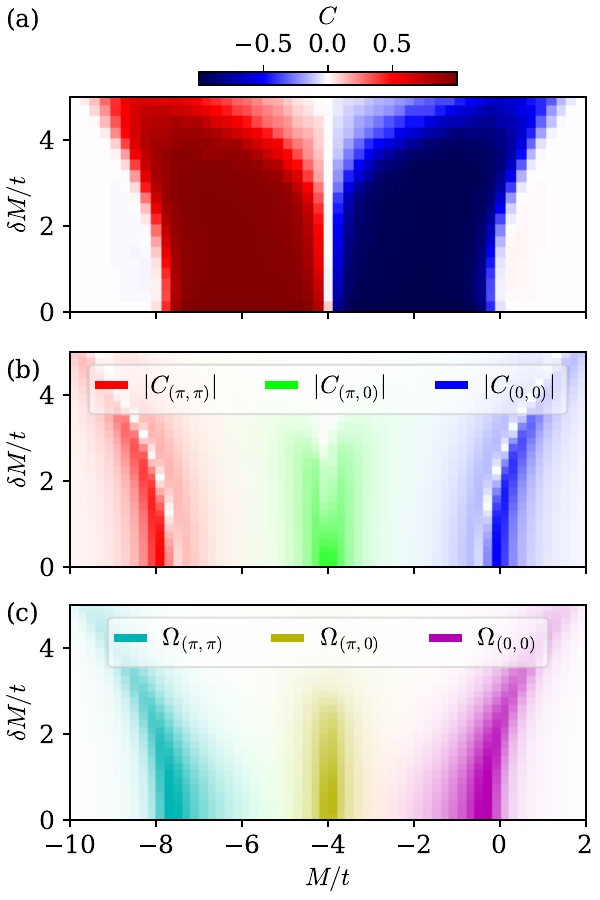}
    \caption{(a) Topological phase diagram of the crystalline model as a function of $M/t$ and disorder strength $\delta M/t$. (b) Topological markers $|C_\bk|$ for $\bk = (0,0)$ (red), $\bk = (\pi,0)$ (green) and $\bk = (\pi,\pi)$ (in blue), as a function of $M/t$ and disorder strength $\delta M/t$. (c) Markers of nonlocal quantum metric $\Omega_\bk$ for $\bk = (0,0)$ (pink), $\bk = (\pi, 0)$ (gold) and $(\pi,\pi)$ (green). In (b,c), color intensity represents the absolute value of the indicator, each with their own hue.}
    \label{fig:signatures}
\end{figure}

Moreover, we show that the patterns uncovered in Fig.~\ref{fig:crystal} can be used as a signature for each topological phase transition. To this end, we introduce integrated markers of the nonlocal Chern number
\begin{equation}
    C_\bk = \iint\diff \br_1\diff \br_2C(\br_1, \br_2) e^{i\bk\cdot(\br_1-\br_2)},\label{eq:indicator}
\end{equation}
and similarly for $\Omega_\bk$, at $\bk = (0,0), (0,\pi)$, and $(\pi,\pi)$.
The exponential phase factor in Eq.~\eqref{eq:indicator} probes a single modulation period, thus allowing us to quantify the weight of the corresponding pattern in the spatial variations of $C(r,r_0)$ and $\Omega(r, r_0)$.
In the absence of disorder ($\delta M = 0$), Fig.~\ref{fig:signatures}(b-c) shows that each marker peaks around a single topological phase transition, exactly where the Chern number $C$ changes, see Fig.~\ref{fig:signatures}(a). When local disorder $\delta M$ is added, the topological phase transition shifts with $M$.
Remarkably, Fig.~\ref{fig:signatures}(b-c) show that the integrated topological markers exactly follow the phase transitions, even for large disorder.
The fact that the hues also do not change indicates that the indicators do not mix with increasing disorder.
Note that further summing Eq.~\eqref{eq:indicator} over all $\bk$ generates the Chern number Eq.~\eqref{eq:C} and quantum metric Eq.~\eqref{eq:omega}, but that misses the important finer structure.
Thus, the real-space pattern of the non-local quantum geometry clearly differentiates topological phase transitions, even between different phase transitions within an AZ class and with disorder present.

\customSection{Diagnosis of amorphous topological phases}
Amorphous systems are fundamentally different from disordered crystalline models: in amorphous materials, momentum is never a good quantum number for analyzing topology because long-range lattice order is completely absent, rather than merely disrupted by disorder.
Thus, the spatially resolved quantum geometric tensor remains the only available tool to identify topological phase transitions. 
Here we show that the nonlocal signatures, identified in Fig.~\ref{fig:crystal}(c) for a crystalline sample, subsist in the amorphous case. We are able to relate this to the remnant local order, especially the local coordination which is dictated by chemistry and thereby generically present in amorphous materials. 

\begin{figure}
    \centering
    \includegraphics[width=\linewidth]{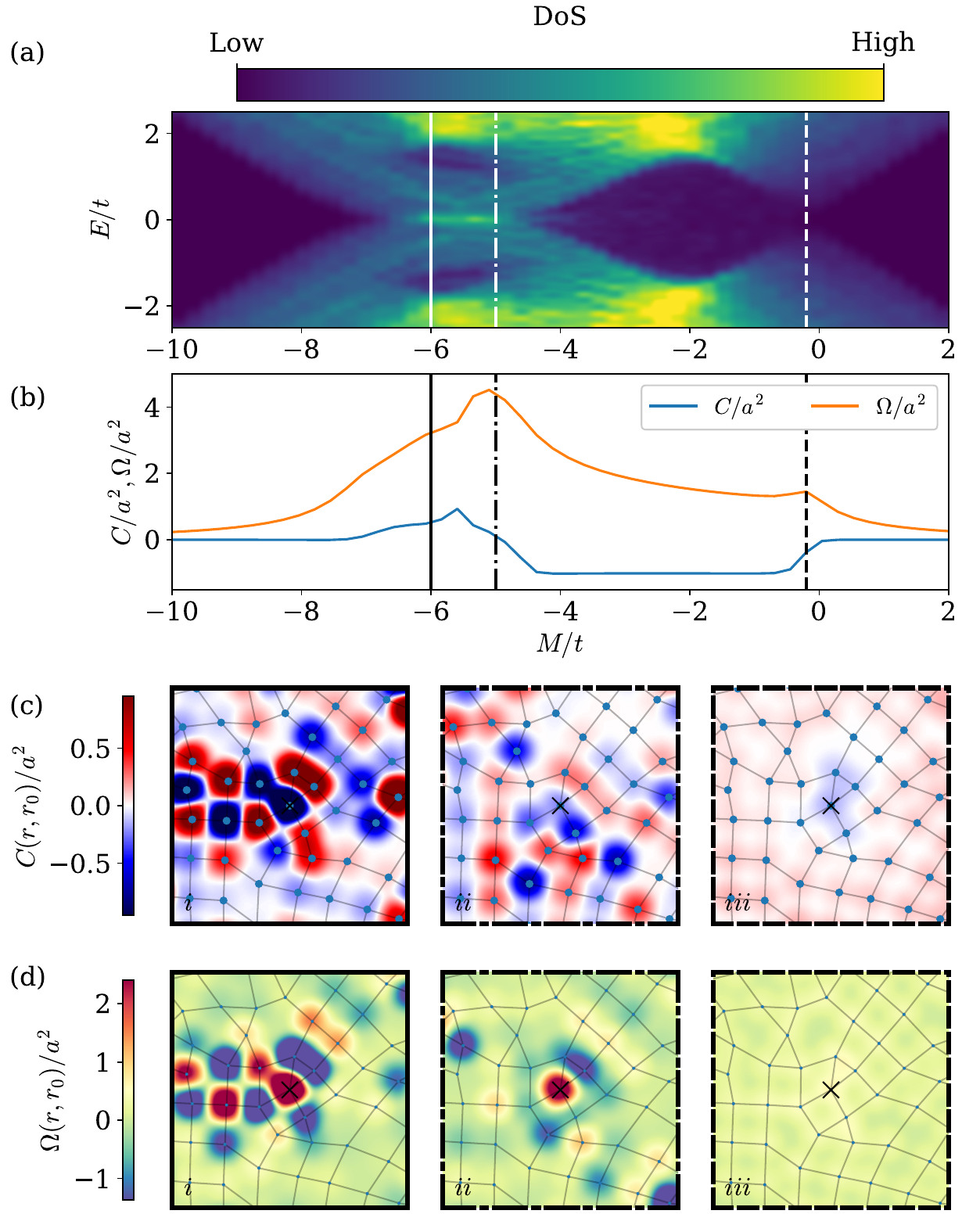}
    \caption{Nonlocal quantum geometry at the topological phase transitions for the amorphous system. 
    (a) Energy spectrum as a function of $M/t$. Compared to the crystalline case (Fig.~\ref{fig:crystal}), the two topological transitions at $M = -8$ and $M=-4$ are merged.
    (b) Local Chern marker $C$ and quantum metric $\Omega$ averaged over the central region.
    (c-d) Real-space patterns of the nonlocal Chern marker $C(r,r_0)$ (c) and quantum metric $\Omega(r,r_0)$ (d) close to each gap closing: ($i$) $M/t = -6$, ($ii$) $M/t = -5$, and ($iii$) $M/t = -0.2$. Solid, dashed, and dashed-dotted vertical lines and frame contours indicate the different $M$ values. Amorphous lattice is drawn for reference and black cross indicates reference site $r_0$.
    }
    \label{fig:amorphous}
\end{figure}

The energy spectrum and quantum geometric markers for the  amorphous version of Hamiltonian Eq.~\eqref{eq:H} are summarized in Fig.~\ref{fig:amorphous}.
A striking difference appears already in the DoS of the crystalline (Fig.~\ref{fig:crystal}(a)) and amorphous (Fig.~\ref{fig:amorphous}(a)) systems: the latter only exhibits two gap closings and thus at most one topological phase. 
This is confirmed by the local Chern marker and local quantum metric in Fig.~\ref{fig:amorphous}(b). The latter still diverges at all gap closings, while the former only shows one topological phase, with $C = -1$. Intriguingly, the phase with $C = 1$ in the crystalline lattice has now disappeared, replaced by a wide metallic region with a non-quantized value of the Chern marker. 
This is surprising as the two phases $C =\pm1$ belong to the same symmetry class A of the AZ classification~\cite{altland1997}, and naively we would expect that they are both equally robust (or not) to amorphousness.
We uncover the reason why one topological phase is much more robust than the other to amorphization in the patterns formed by the nonlocal quantum geometry in position space, see Figure~\ref{fig:amorphous}(c-d).
Each panel shows how $C(r,r_0)$, in Fig.~\ref{fig:amorphous}(c) and $\Omega(r, r_0)$, in Fig.~\ref{fig:amorphous}(d) depend on $r$ relative to a given site at position $r_0$ in the center, marked by a cross. 
We again use $M$ values where the local quantum metric peaks.

It is first interesting to note that we somewhat recover real-space patterns of the quantum geometry similar to those observed in the crystalline case (see Fig.~\ref{fig:crystal}(c-d)).
This is particularly apparent for the transition at $M = 0$ that seems completely unaffected by the amorphousness. 
We can relate this robustness to the real-space pattern. 
It shows no sign change in the close environment of the site $r_0$ in Fig.~\ref{fig:crystal}(c). 
Thus, it is compatible with any polygon that nearest neighbor atoms may form in a lattice, as also seen in the ($iii$) panels of Fig.~\ref{fig:amorphous}(c-d) that keeps a similar local uniformness. 
This could equivalently be anticipated from the momentum space picture: the gap closing of the $M = 0$ transition occurs at $\bk a = (0,0)$, a Bloch momentum with no spatial modulation.

Between $M/t \sim -5$ and $M/t\sim -6$, the amorphous system shows an extended gapless phase, with no crystalline equivalent. 
Across this metallic phase, we identify large plateau in the local quantum metric $\Omega(r_0, r_0)$, as expected for a metallic state.
Plotting the nonlocal markers $C(r, r_0)$ and $\Omega(r, r_0)$ in Fig.~\ref{fig:amorphous}(c-d) sheds light on its origin.  
We find a deformed cross-square pattern close to $M/t = -5$, and a deformed checkerboard pattern close to $M/t = -6$, i.e., deformed versions of Fig.~\ref{fig:crystal}(c-d).
Still, the topological phase transitions taking place at $(\pi,\pi)$ and $(0,\pi)$ in the crystal is gone in the amorphous system, thus shrinking out the topological phase $C = 1$. We can explain this readily from the nonlocal quantum geometry: in the crystalline sample, the cross-square and checkerboard patterns differ by the value they take on second nearest neighbor sites. 
In the amorphous lattice, sites with a square local environments have well-defined second-nearest neighbors, thus exhibiting a similar quantum geometry pattern as the crystal. 
However, amorphous systems also naturally host triangles among the polygons formed by the sites. Then, some sites are at the same time be both nearest and second-nearest neighbors. This results in a mixing of the two patterns beyond nearest neighbor sites and consequently the destruction of the crystalline $C=1$ phase, instead creating an extended metallic region, where the quantization of the Chern number is lost.
Our results are again consistent with the momentum space picture: the crystalline topological phase transitions at $M/t = -8$ and $M/t = -4$ occur due to gap closings at different points along the edge of the Brillouin zone. 
However, the lattice disorder makes the amorphous system isotropic, meaning all points lying at equal distance from $\bk=0$ are equivalent. Thus, approximately, all points on the Brillouin zone edge behave similarly and produce rather similar nonlocal signatures for these $M/t$ values.
These results showcase that the nonlocal quantum geometry is a powerful tool to explain why certain amorphous topological phases remain more robust than others.

\customSection{Conclusion and Discussion}
We show that the nonlocal component of spatially resolved quantum geometric markers, both the Chern and quantum metric markers, provides a robust diagnostic of topological phase transitions, even in the presence of disorder and amorphousness. 
Their characteristic spatial patterns serve as clear signatures of transitions and provide an intuitive understanding of the stability, or fragility, of topological phases in imperfect crystals as well as fully amorphous systems. This demonstrates their utility as a unifying framework for probing the stability of topological phases across ordered and amorphous materials, in particular, answering why some phases exhibit markedly different stability even within the same AZ symmetry class.

Looking ahead, our results suggest that there are several promising future directions. 
First, generalizing this classification by relating the spatial symmetry of the nonlocal topological markers to those of the atomic orbitals at play in different topological systems, including in higher dimensions, would provide a systematic approach to predict topological phases in disordered and amorphous systems, similar to topological quantum chemistry~\cite{bradlyn2017}. 
Second, the real space patterns of nonlocal markers may be related to Berry curvature dipoles, thereby offering an intriguing pathway to nonlinear Hall effects and their response functions~\cite{sodemann2015, ortix2021, bhattacharya2025} also in amorphous samples. 
Overall, the connection we establish between position and momentum space perspectives provides a general strategy for probing more complex and realistic topological scenarios, where conventional momentum space diagnostics fail. 
As such, our work positions nonlocal, spatially resolved topological markers as a fresh tool for diagnosing topological phase transitions, offering both new conceptual insight and practical guidance for the exploration of disordered and amorphous materials.

\customSection{Acknowledgements}
We thank U.~Nitzsche for technical assistance, and R.~Arouca, A.~Bhattacharya and A.~G.~Grushin for interesting and fruitful discussions related to this work. 
All authors acknowledge support from the Knut and Alice Wallenberg Foundation through the project grant 2019.0068.
Q.M.~and A.B.-S.~acknowledge funding from the European Research Council (ERC) under the European Union's Horizon 2020 research and innovation programme (ERC-2022-CoG, Grant agreement No.~101087096).
H.L.~and E.J.B.~were supported by the Swedish Research Council (VR, grant 2024-04567), the Wallenberg Scholars program of the Knut and Alice Wallenberg Foundation (2023.0256), and the G\"oran Gustafsson Foundation for Research in Natural Sciences and Medicine.
Our calculations were performed using the Python package kwant~\cite{groth2014} and our plots using matplotlib~\cite{hunter2007}. The code used for the numerical calculations and the data shown in the manuscript are available at Ref.~\cite{zenodo}.

\bibliographystyle{apsrev4-1}

\bibliography{bibliography}

\end{document}